# Non-adiabatic Modal Dynamics around Exceptional Points in an All-Lossy Dual-Mode Optical Waveguide: Towards Chirality Driven Asymmetric Mode-Conversion

Arnab Laha,[1] Abhijit Biswas,[2] and Somnath Ghosh [1, *]

[1] Department of Physics, Indian Institute of Technology Jodhpur, Rajasthan-342011, India
[2] Institute of Radiophysics and Electronics, University of Calcutta, Kolkata-700009, India
* *somiit@rediffmail.com*

We report a 1D planar optical waveguide with transverse distribution of inhomogeneous loss profile, which exhibits an exceptional point (EP). The waveguide hosts two leaky resonant modes; where the interaction between them in the vicinity of the EP is controlled by proper adjustment of the inhomogeneity in attenuation profile only. We study the adiabatic dynamics of propagation constants of the coupled modes by quasi-static encirclement of control parameters around the EP. Realizing such an encirclement with the inhomogeneous loss distribution along the direction of light propagation, we report the breakdown of adiabatic evolution of two coupled modes through the waveguide in presence of an EP. Here, during conversion the output mode is irrespective of the choice of input excited mode but depends on the direction of light transportation. This topologically controlled, robust scheme of asymmetric mode conversion in the platform of the proposed all-lossy waveguide structure may open up an extensive way-out for implementation of state-transfer applications in chirality driven waveguide-based devices.

## 1. Introduction:

Over the past decade, apart from the familiar Hermitian quantum systems, realistic open systems with richer physical impacts have attracted considerable attention as they interact with the environment. To describe these open systems, non-Hermitian quantum mechanics can efficiently be employed because this formalism deals with some systematic tools which are able to reduce a physical system into an effective Hamiltonian [1, 2]. Recently, the enticing features of non-Hermitian quantum mechanics enrich the platform for topological study on various quantum-inspired or wave-based photonic structures. Contextually, one of the contemporary astonishing topological feature of such photonic systems is the appearance of hidden singularities, namely exceptional points (EPs) [2]. An EP can be defined as a particular point in 2D (at least) system parameter space where the coupled eigenvalues and the corresponding eigenstates of the underlying Hamiltonian simultaneously coalesce, and are connected by a second order branch point. Such existence of an EP can effectively be understood by the phenomenon of avoided resonance crossing (ARC) between two respective interacting eigenvalues with crossing/anticrossing of their frequencies and lifetimes [2, 3]. Thus, an EP can be considered as a special type of degeneracy which is characteristically far away from the conventional Hermitian degeneracies (with same eigenvalues but different eigenfunctions) like diabolic points, Dirac points, etc.

The fascinating features of EPs have extensively been explored in optical waveguides [4–7], microcavities [8], lasers [9, 10], photonic crystals [11, 12], etc. and also in several non-optical systems like microwave billiards [13], atomic [14] and molecular systems [15], etc. The topological structure of an EP has first experimentally been demonstrated in a microwave cavity [16] with an explicit observation of a chiral behavior [17]. However, apart from the non-optical structures, widely available photonic systems with precise fabrication control can provide an highly promising fertile domain to study the fundamental aspects of EPs with several enticing technological applications; just for example, asymmetric mode conversion/switching [4–7], unidirectional light refection or transmission [18], resonance assisted tunneling [19], dark state lasing [20] ultra-sensitive EP-aided sensing [21–23], optical non-reciprocity [24], to name a few.

The complex eigenvalues unveil non-trivial topological behavior alongside an EP; which can be explored by encircling the EP with quasi-static variation of system parameters along a closed loop. This EP encircling scheme play a key role behind the non-Hermitian phenomena of flip-of-states [8, 14] and also asymmetric mode-conversion/switching [4–7]. Such a parametric encirclement allows the continuous swapping between the coupled eigenvalues in the complex eigenvalue plane [3]. For one complete encirclement, the coupled eigenvalues exchange their initial positions. Simultaneously, one of the underlying eigenstate experiences a phase change i.e., $\{\Psi_1, \Psi_2\}$(one round) $\rightarrow \{\Psi_2, -\Psi_1\}$ or $\{-\Psi_2, \Psi_1\}$ [25]. Here, an EP acts as a $2^{nd}$-order branch point for eigenvalues and a $4^{th}$-order branch point for the eigenvectors.

Now, when an EP is slowly encircled in parameter space, the permutation between the complex eigenvalues are pursuing adiabatic evolution. However, when we consider a dynamical encirclement i.e., the system parameters vary in time, the adiabaticity in dynamics of the system breaks down and one of the state from corresponding coupled pair behaves non-adiabatically [26, 27]. This yields chiral transportation of light; where at the output it is converted in a specific mode, irrespective of the choice of input excited mode. Here, a clockwise and an anticlockwise dynamical parametric encirclement around EP will lead different final states beyond the adiabatic restrictions [28].

An EP directly deals with the system non-Hermiticity which is realized in terms of gain or/and loss, etc. While a system operating in the vicinity of an EP, gain-loss can remarkably affect its phenomenal features. Here, an EP can be reached and also manipulated with proper tuning of the amount of gain-loss. For an EP which is effectively encircled with adiabatic variation of system gain-loss, the EP-aided flip-of-states phenomenon has reported in the contexts of partially pumped optical microcavity [8], photonic crystal [11], etc.; whereas the effect of dynamical encirclement of an EP, exploiting beam propagation in a dual mode optical waveguide with proper gain-loss modulation, has been predicted in a dielectric waveguide by Ghosh et. al. [7] and experimentally been reported in a metallic waveguide by Doppler et. al. [4]. However, it should be quite interesting for a real system and also fabrication feasible, if such properties of an EP can be achieved by patterning the loss profile only, rather than simultaneous presence of both gain and loss. To introduce gain in a system, one should consider active pumping from external sources in terms of perturbations which may assemble considerable noise in the system; where increasing gain or decreasing gain beyond a particular threshold can cause the system to become unstable. Moreover, in a gain -induced system, gain itself can guide a mode. This gain guided mode may also be able to interact with the actual scattering modes of the system and demolish the desired performance of the system. Thus, if we design a gain-free all-lossy structure then the described limitations can be overcome; where the interaction between two scattering modes can be controlled by system loss profile only. Such an all-lossy photonic structure is yet to be explored and also not been studied in the context of EP-aided asymmetric mode conversion. This scheme should be accessible for the majority of conventional optical elements and more feasible for device realization.

In this article, we report a step-index symmetric planar dual mode optical waveguide. Here, non-hermiticity is attained with distribution of an inhomogeneous loss profile along the transverse direction; where in the core region fractional loss ratio between two halves is properly maintained. The waveguide hosts a second order EP which is manipulated by the controlled interactions between two supported modes with proper tuning of the attenuation-profile only. Such an all-lossy waveguide structure, operating at an EP, is reported for the first time to the best of our knowledge; where no external gain is present. Here, the EP is dynamically encircled by longitudinal variation of loss profile with simultaneous maintenance of loss-inhomogeneity in the waveguide. Following this dynamical parametric encirclement, we study the propagation of two supported modes and present a chirality driven asymmetric mode conversion scheme; where we observe the individual conversion from both the modes into one specific mode, regardless of the choice of excited mode but depending on the direction along which the EP is encircled. We measure the efficiencies for all the conversions. To take into account the fabrication tolerances, the immutability of the asymmetric mode conversion scheme is also studied against small spatial fluctuations in refractive index modulation within paraxial limit. Our proposed all-lossy waveguide structure with specific chirality-driven mode conversion scheme paves an efficient platform to develop optical switches, converters, etc.

## 2. OPTICAL WAVEGUIDE HOSTING AN EP:

The scenario of the appearance of an EP in a dual mode optical waveguide can conveniently be illustrated by considering a $2 \times 2$ non-Hermitian Hamiltonian as

$$H = \begin{pmatrix} \beta_m + i\gamma_m & \kappa \\ \kappa^* & \beta_n + i\gamma_n \end{pmatrix} \tag{1}$$

Here, $\beta_m$ and $\beta_n$ represent the real propagation constants of two consecutive supported modes with respective losses $\gamma_m$ and $\gamma_n$. $\kappa$ is a coupling term ($\kappa^*$ indicates its complex conjugate). Introducing complex propagation constants $\tilde{\beta}_j = \beta_j + i\gamma_j$; ($j = m, n$), the eigenvalues of the Hamiltonian H can be written as

$$\beta_\pm = \frac{\tilde{\beta}_m + \tilde{\beta}_n}{2} \pm \sqrt{\left(\frac{\tilde{\beta}_m - \tilde{\beta}_n}{2}\right)^2 + |\kappa|^2} \equiv \tilde{\beta}_{av} \pm \sqrt{\left(\delta\tilde{\beta}\right)^2 + |\kappa|^2} \; ; \tag{2}$$

with

$$\tilde{\beta}_{av} = \frac{\tilde{\beta}_m + \tilde{\beta}_n}{2} \; , \; \delta\tilde{\beta} = \frac{\tilde{\beta}_m - \tilde{\beta}_n}{2}. \tag{3}$$

An EP refers to a particular branch point where two coupled eigenvalues coalesce, i.e. $\beta_+ = \beta_-$. Thus in complex $\kappa$-plane, EP appears at $\kappa_{EP} = \pm i\delta\tilde{\beta}_{EP}$. Here, the location of an EP can be entirely accessed by real and complex parts of $\delta\tilde{\beta}$ (where, $\delta\tilde{\beta} \equiv \delta\beta + i\delta\gamma$) with fulfillment of the conditions as $\beta_m = \beta_n$ and $\kappa = |\gamma_m - \gamma_n|/2$.

## 3. DESIGN OF THE OPTICAL WAVEGUIDE:

### A. Waveguide specifications with operating parameters-

Analogically, to implement the idea as described in the Sec. II, we design an optical waveguide which is schematically shown in Fig. 1. We consider propagation along the $z$-direction with respect to the transverse $x$-direction. The waveguide, occupying the region $-W/2 \leq x \leq W/2$, consists a core with passive refractive index $n_h$, which is surrounded by a cladding having the same as $n_l$.

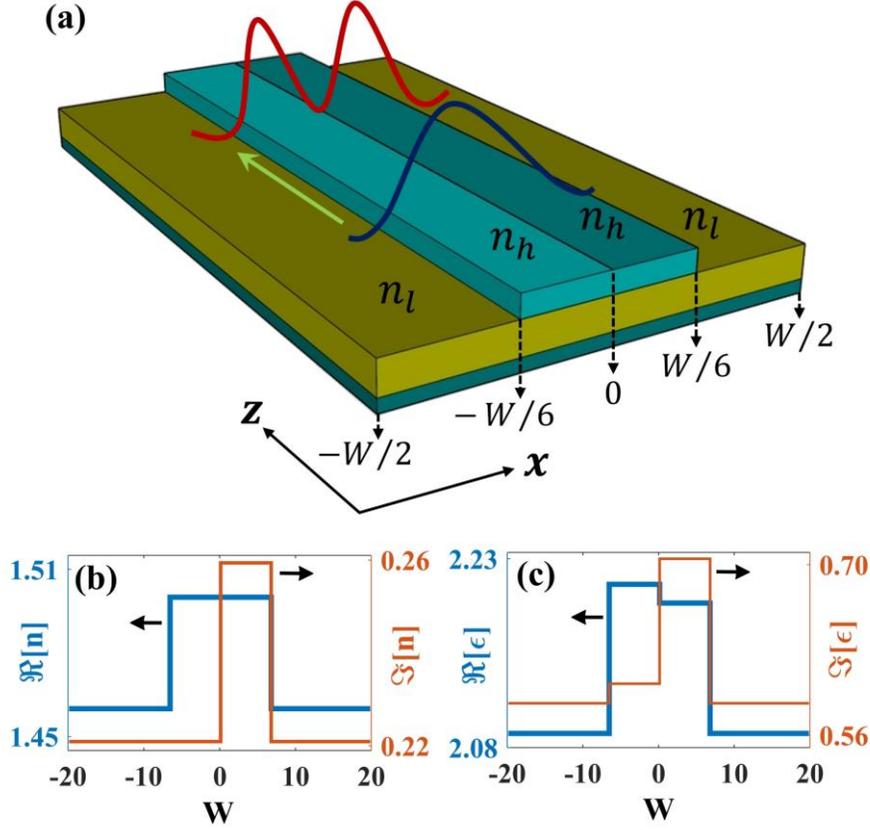

**FIG. 1.** (Color online) **(a)** Schematic of the proposed dual-mode optical waveguide that occupies the region $W/2 \leq x \leq W/2$ with $W = 40$ (dimensionless). The propagation is considered along the length ($z$ axis). The core and cladding refractive indexes are chosen as $n_h = 1.50$ and $n_l = 1.46$ respectively. **(b)** Transverse complex refractive index profile $n(x)$. Blue curve shows the $\Re(n)$ and the brown curve shows the $\Im(n)$ while operating at an EP. **(c)** Transverse distribution of relative permittivity $\epsilon x)$, where $\epsilon = n^2$.

Here the non-hermiticity is introduced by specific transverse distribution of the inhomogeneous loss profile, with loss-coefficient $\gamma$. Inhomogeneity in loss-profile is maintained by a parameter $\tau$ which essentially denotes the fractional-loss-ratio. Such loss distribution can be realized by spatially patterned imaginary part of the refractive indices. Thus, under operating condition, the transverse distribution of $n(x)$ can be given as follows.

$$n(x) = \begin{cases} n_h + i\gamma, & -\dfrac{W}{6} \leq x \leq 0 \\ n_h + i\tau\gamma, & 0 \leq x \leq \dfrac{W}{6} \\ n_l + i\gamma, & \dfrac{W}{6} \leq |x| \leq \dfrac{W}{2} \end{cases} \quad (3)$$

This complex refractive index profile $n(x)$ and corresponding profile of the relative permittivity $\epsilon(x)$ are shown in Fig. 1(b) and (c), respectively. Due to presence of such inhomogeneous loss profile, the resulting waveguide does not follow the *PT*-symmetric constraints. Here, it is evident that a specific loss distribution is patterned along the transverse direction which makes the waveguide all-lossy, i.e., no external gain is present. The system allows spatial loss-variation along the propagation direction, i.e., along the length scale only; which depends on distinct tunabilities of the control parameters $\gamma$ and $\tau$. However, for each cross section of the waveguide the loss distribution is fixed as described in Eq. 3. Here, obeying causality condition, the independent tunability of $\Im(n)$ along the length-axis irrespective of the choice of $\Re(n)$ is realized only at single operating frequency; which is dictated by the Kramers-Kronig relation [29].

During operation, we normalize the operating frequency $\omega = 1$ (i.e. the free-space wavelength $\lambda = 2\pi$) and set the total width of the waveguide $W = 40$ in a dimensionless unit; where $W = 20\lambda/\pi$. Here, conveniently we choose μm unit for realistic illustration. The real refractive indices are chosen as $n_h = 1.50$ and $n_l = 1.46$. Such a prototype can suitably be realized by thin-film deposition of glass material over a thick silica glass substrate. These operating parameters are specifically chosen to make sure that the waveguide supports only two quasi-guided modes i.e., the fundamental mode ($\psi_0$) and first-higher-order mode ($\psi_1$). Here the values of propagation constant ($\beta$) are calculated by solving the scalar modal equation corresponding to a steady-state mode-profile $\psi(x)$ given as

$$[\partial_x^2 + n^2(x)\omega^2 - \beta^2]\psi(x) = 0 \quad (4)$$

As the index difference between the core and cladding regions is small, the above scaler equation (given by Eq. 4), derived from the Maxwell's equations, is justified as long as the leaky modes are considered.

## B. Encounter of an exceptional point-

To encounter an EP in our numerically designed waveguide, the concept of avoided resonance crossing (ARC) is exploited [3, 7, 8]. Here, the complex propagation constants ($\beta$) of the pair of coupled modes exhibit special ARCs with crossing/ anti-crossing of their real and imaginary parts, once we vary the loss-coefficient $\gamma$. Now, for different values of fractional-loss-ratio $\tau$, the evolutions of $\Re[\beta]$ and $\Im[\beta]$ with respect to $\gamma$ are recorded in Fig. 2. Red and blue curves represent the propagation constants of $\psi_0$ and $\psi_1$, say $\beta_0$ and $\beta_1$, respectively.

At first, we set $\tau = 1.165$ and tune $\gamma$ extremely slowly from 0 to 0.3. In this situation, the propagation constants exhibit ARC in complex $\beta$-plane with gradual increase in $\gamma$; where $\Re[\beta]$ undergoes an anti-crossing with a simultaneous crossing in $\Im[\beta]$ as shown in Fig. 2(a.1) and (a.2), respectively. Now, while we slightly increase $\tau = 1.175$, a different kind of ARC between $\beta_0$ and $\beta_1$ is observed as can be seen in Fig. 2(b.1) and (b.2), where $\Re[\beta]$ experiences a crossing and $\Im[\beta]$ undergoes an anti-crossing, respectively. The sudden transition between two topologically variant behaviors of ARCs for two different $\tau$- values as depicted in Fig. 2(a) and (b), clearly indicate the appearance of a square root branch point where two coupled states are analytically connected [7, 8].

Now, to locate the approximate position of this branch point, we judiciously choose an intermediate $\tau = 1.17$ and track the dynamics of the complex $\beta$s in Fig. 2(c) with gradual variation of $\gamma$ within the same range. Here, it is clearly noticeable that when $\gamma$ takes the value 0.2008, $\beta_0$ and $\beta_1$ coalesce and loose their identities; which certainly indicates an EP. Hence, in the ($\gamma, \tau$)-plane we numerically locate the EP at $\gamma_{EP} = 0.2008, \tau_{EP} = 1.17$.

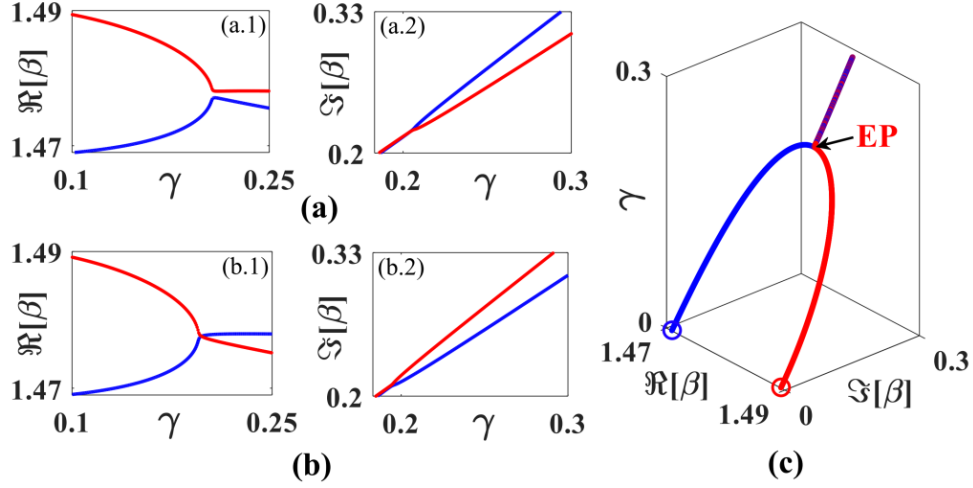

**FIG. 2.** (Color online) **(a)** Dynamics of $\beta_0$ and $\beta_1$ (denoted by red and blue dotted curves) exhibiting ARC with increase in $\gamma$ for a fixed $\tau = 1.165$ with (a.1) anti-crossing in $\Re[\beta]$ and (a.2) crossing in $\Im[\beta]$. **(b)** Similar trajectories for $\tau = 1.175$ with (b.1) crossing in $\Re[\beta]$ and (b.2) anti-crossing in $\Im[\beta]$. **(c)** Trajectories of complex $\beta$-s with respect to $\gamma$ for an intermediate $\tau = 1.17$; where $\beta_0$ and $\beta_1$ coalesce near $\gamma = 0.2008$. The big circular markers with respective colors represent their initial positions, i.e. at $\gamma = 0$.

## 4. PROPAGATION CHARACTERISTICS OF THE WAVEGUIDE IN PRESENCE OF AN EP:

### A. Modal analysis around the EP: Flip-of-states-

In this section, we study the effect of encirclement around the identified EP in the 2D parameter space; where the parameters $\gamma$ and $\tau$ are varied in a specific way. Here, a closed circular contour is considered with center at the EP, i.e. at ($\gamma_{EP}$, $\tau_{EP}$) following the coupled equations-

$$\gamma(\phi) = \gamma_{EP}[1 + \eta \cos \phi], \quad \tau(\phi) = \tau_{EP}[1 + \eta \sin \phi] \tag{5}$$

Here, $\eta$ ($\in (0, 1]$) represents a characteristics parameter which is equivalent to the radius, and the tunable angle $\varphi$ ($\in [0, 2\pi]$) essentially acts as a control knob of adiabaticity in variation of $\gamma$ and $\tau$ along the closed contour. This encircling method opens a route to check the singular behavior of the identified EP by scanning a large area around it.

Fig. 3(a) represents such a closed loop around the EP (marked by red cross) in the ($\gamma, \tau$)-plane with $\eta = 0.1$. In Fig. 3(b), we study the effect of this parametric encirclement on the dynamics of the propagation constants of two corresponding coupled modes in the complex $\beta$- plane. With enough small steps (almost quasi-statically) on the enclosing loop along the anticlockwise direction, the adiabatic motions of $\beta_0$ and $\beta_1$ are properly traced. Each point on red and blue trajectories in Fig. 3(b) represents the evolution of $\beta_0$ and $\beta_1$ from their respective initial positions (when $\phi = 0$) which correspond to each point of evolution on brown parametric loop in Fig. 3(a). Interestingly, one round encirclement in the ($\gamma, \tau$)-plane around the EP results the permutation between $\beta_0$ and $\beta_1$, i.e. they are mutually exchanging their positions in the complex $\beta$-plane and form a complete loop. However, following the next round along the same contour they regain their initial positions; serving EP as second order branch point for eigenvalues. In the Fig. 3(c) and (d), the individual trajectories of $\Re[\beta]$ and $\Im[\beta]$ are shown with simultaneous variation of $\gamma$ and $\tau$ along the closed loop as described in Fig. 3(a), respectively. Arrows indicate the direction of progression in both the $\beta$-plane and $(\gamma, \tau)$-plane. Such a phenomenon in complex $\beta$-plane can be referred as flip-of-states and more specifically, $\beta$- *switching*. On the contrary, it should be also observed that if the EP is not properly enclosed in the parameter plane the propagation constants make individual loops in $\beta$-plane instead of permutation.

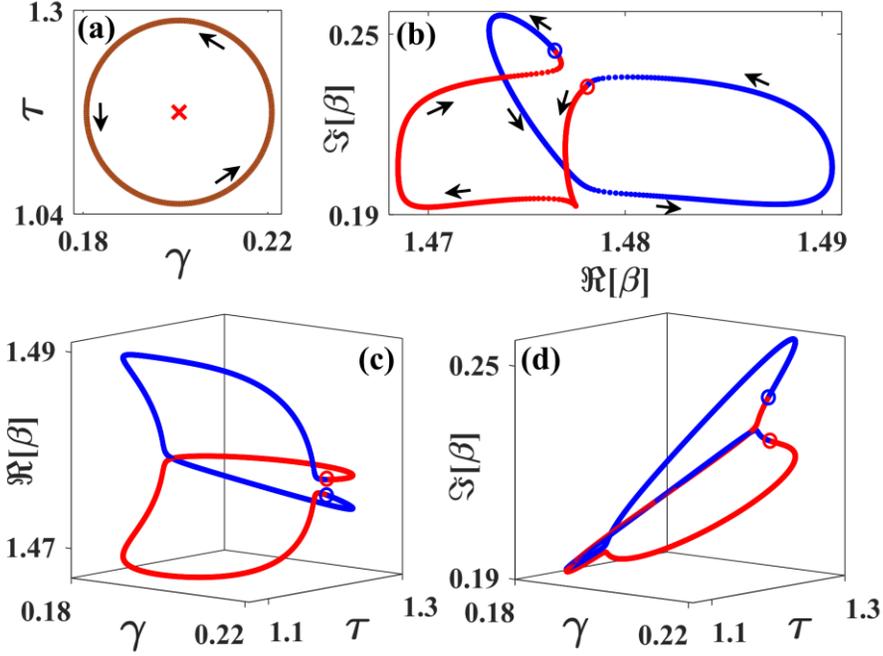

**FIG. 3.** (Color online) **(a)** Anticlockwise encircling process around the EP in the $(\gamma, \tau)$-plane as defined in Eqs. 5 with center at $\gamma_{EP} = 0.2008$, $\tau_{EP} = 1.17$ (indicated by the red cross), and $\eta = 0.1$ **(b)** Trajectories of $\beta_0$ and $\beta_1$ represented by red and blue dotted curves for one round encirclement as described in (a); along-with individual dynamics of **(c)** $\Re[\beta]$ and **(d)** $\Im[\beta]$ from their respective initial positions (i.e. when $\phi = 0$) as indicated by circular markers of respective colors. Arrows indicate the direction of progressions.

A leading drawback of the described EP encircling method, while we choose a circle as an enclosing loop (Eqs. 5), is that one can never reach the situation when $\gamma = 0$. Even at $\phi = 0$, the coupled modes experience some amount of losses. Thus, before permutation the initial positions of $\beta_0$ and $\beta_1$ are shifted from their actual passive locations (i.e. the positions when $\gamma = 0$) in the complex $\beta$-plane. This is a crucial requirement for any device applications that one should start and end at passive system after encirclement. This demands an enclosing loop in he $(\gamma, \tau)$-plane which consider the situation $\gamma = 0$ at both the input and output. To ensure this, we choose a different closed contour around the identified EP by replacing the Eqs. 5 with

$$\gamma(\phi) = \gamma_0 \sin\left(\frac{\phi}{2}\right); \; \gamma_0 > \gamma_{EP}, \quad \tau(\phi) = \tau_{EP} - \eta \sin\phi \tag{5}$$

Eqs. 6 open a way to achieve passive modes avoiding any loss-dominated modes at the input and output interfaces. Here, the clockwise operation is carried out by choosing $\eta < 0$ whereas for anticlockwise operation, we have to set $\eta > 0$. Note that, if $\gamma_0 < \gamma_{EP}$, the parametric loop does not enclose the EP. Now, choosing $\gamma_0 = 0.25 \, (> \gamma_{EP})$ and $\eta = 0.05$, we describe a clockwise variation of $\gamma$ and $\tau$ along a closed contour in Fig. 4(a) which encloses the embedded EP properly. In Fig. 4(b), we record the corresponding movements of $\beta_0$ and $\beta_1$ from their respective passive positions (i.e., where $\Im[\beta] = 0$ for both of them) in the complex $\beta$-plane for a complete parametric loop ($0 \leq \varphi \leq 2\pi$) in the $(\gamma, \tau)$-plane. Here also the propagation constants switch their positions at the end of the encircling process in a generic fashion as described previously in Fig. 3 (for circular loop); however, in this case, they exactly exchange their passive locations in the $\beta$-plane.

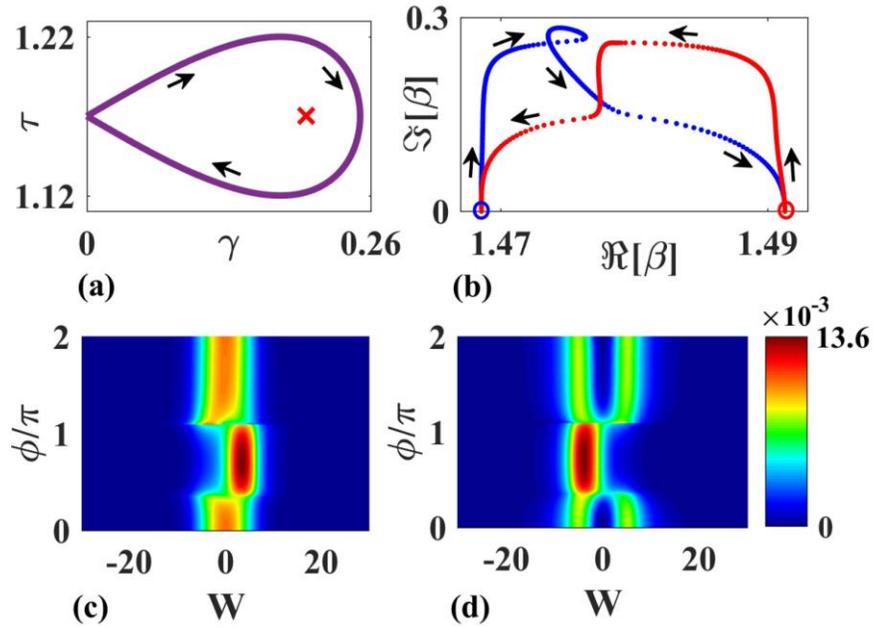

**FIG. 4.** (Color online) **(a)** A clockwise variation of $\gamma$ and $\tau$ around the EP (indicated by the red cross) along the loop defined by Eqs. 6 with $\gamma_0 = 0.25 \, (> \gamma EP)$ and $\eta = 0.05$. **(b)** Corresponding dynamics of $\beta_0$ and $\beta_1$ represented by red and blue dotted curves for one round operation. **(c, d)** Evolution of (c) $\psi_0$ and (d) $\psi_1$ (mode steaking) around the EP for each set of $(\gamma, \tau)$ on the loop described in (a).

In the Fig. 4(c) and (d) we plot the modal intensities $|\psi(x)|^2$ (by color axis) of the underlying eigenmodes $\psi_0$ and $\psi_1$ respectively for each value of $\phi$, i.e., for each set of $(\gamma, \tau)$ on the parametric loop during one complete clockwise encirclement as described in Fig. 4(a) (Note that, this is not the propagation of the eigenmodes along the waveguide; which will be studied in the following sections using beam propagation simulation method, this is mode stacking). Now, from both Fig. 4(c) and (d), it is evident that both the eigenstates lose their identities around the EP and approximately near EP corresponding $|\psi(x)|^2$ pick-up highest magnitudes.

## B. Propagation of modes along the waveguide: Asymmetric mode conversion-

In the preceding section, we analyze the evolutions of the supported modes with simultaneous study of the dynamics of the corresponding propagation constants and then state-flips. Such EP aided state exchange phenomena bargain with adiabatic theorem; where it has been observed that the propagation constants corresponding to the pair of coupled modes are permuted adiabatically if the control parameters vary in extremely small steps. In this context, if we consider two level system having a time dependent Hamiltonian $H(t)$, the evolutions of the corresponding eigenstates are administrated by time dependent Schrödinger equation (TDSE). If the considered system is Hermitian then for extreme slow variation of system parameters, the obtained adiabatic solutions are accompanied with the exact solutions of the TDSE. However, in case of a non-Hermitian system, the adiabatic theorem no-longer well founded essentially due to presence of an EP with associated non-Hermitian components [26, 27].

Here, we present a brief mathematics behind the breakdown in adiabaticity. For instance, we consider the effective non-Hermitian Hamiltonian associated with our designed waveguide as $H(\vec{\mu})$ that depends on a time dependent (length dependent) parameter $\vec{\mu}(t)$. For our waveguide the simultaneous time variations of $\gamma$ and $\tau$ can be analogically compared with the variation of the vector $\vec{\mu}$. Now, the state at time t can be expressed as

$$|\Psi(t)\rangle = \sum_n c_n(t) e^{-i\phi_n t} |n(\vec{\mu}(t))\rangle \qquad (7)$$

with instantaneous eigenvalues $E_n(\vec{\mu}(t))$ and eigenstates $|n(\vec{\mu}(t))\rangle$ at the time $t$. Here,

$$\phi_n(t) = \int_0^t E_n(\vec{\mu}(t'))dt'. \qquad (8)$$

Now the substitutions of Eq. 7 and Eq. 8 into the TDSE results

$$i\sum_n \dot{c}_n e^{-i\phi_n} |n(\vec{\mu})\rangle = -\sum_n c_n e^{-i\phi_n} \dot{\vec{\mu}} \cdot \partial_\mu |n(\vec{\mu})\rangle. \qquad (9)$$

Here we use $\left\langle n \left| \frac{\partial}{\partial t} \right| n \right\rangle = \langle n|\partial_\mu|n\rangle \dot{\vec{\mu}}$ with $\partial_\mu \equiv \frac{\partial}{\partial \mu}$. Now, exploiting the Eq. 9, we can examine the self-consistency of the state $|1\rangle$ at an instantaneous time $t$ with the state $|0\rangle$ as

$$i\dot{c}_0 e^{-i\phi_0} \approx -c_1 e^{-i\phi_1} \dot{\vec{\mu}} \langle 0|\partial_\mu|1\rangle. \qquad (10)$$

If we assume that $|1\rangle$ and $|0\rangle$ are approximately power orthogonal then from Eq. 10,

$$\dot{c}_0 \approx ic_1 \exp\left[-i\int_0^t \left[E_0(\vec{\mu}(t')) - E_1(\vec{\mu}(t'))\right] dt'\right] \dot{\vec{\mu}} \langle 0|\partial_\mu|1\rangle. \qquad (11)$$

Now, for usual Hermitian cases, the exponential term in Eq. 11 indicates just a phase factor. So, for extremely small values of $\dot{\vec{\mu}}$ i.e.; the adiabatic time evolution of the system parameter, $\dot{c}_0$ should be suppressed; where the amplitude $c_1(t)$ should dominate all the other amplitudes for the entire evolution time. This means that the system meets criterion of the adiabatic theorem. However, for our non-Hermitian waveguide we deal with imaginary eigenvalues and corresponding decaying eigenstates. Here, the exponential term in Eq. 11 is not generally a phase factor. If this is a growing exponential, then the amplitude of $|0\rangle$ grows rapidly for extremely small $\dot{\vec{\mu}}$ along the loop in parameter space; where self-consistency of the above calculation is no longer valid. Thus, for the dominant influence of breakdown of adiabaticity in non-Hermitian systems, only the eigenstates having comparably lower loss evolves adiabatically with adiabatic change of the system parameters with time along a closed contour around the EP in the 2D parameter space. Simultaneously, the other eigenstates from the corresponding coupled pair having comparably higher decaying rate does not meet the adiabatic expectations and dramatically returns to itself at the end of the encircling process. Thus, for a dynamically encircled EP, the expected adiabatic state-exchange phenomena cannot be reached even for a valid adiabatic exchange of corresponding eigenvalues as discussed in the preceding sections. Here, the actual state-conversion/switching should be asymmetric.

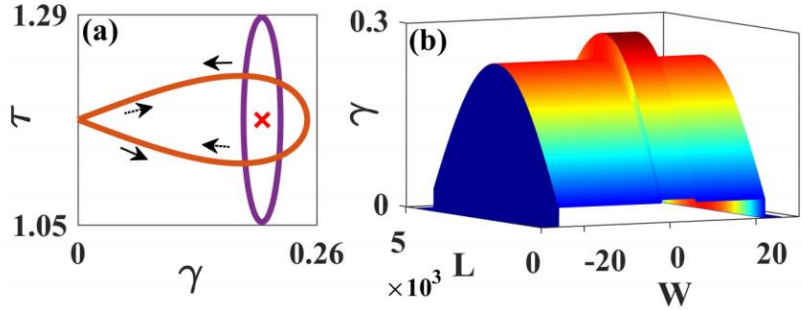

**FIG. 5.** (Color online) **(a)** The parametric loops in the $(\gamma, \tau)$- plane around the EP (indicated by the red cross) for comparison. Here, the brown and violet contours follow Eqs. 5 and Eqs. 6, respectively. Dotted and solid arrows show the clockwise and anti-clockwise directions respectively. **(b)** Loss profile experienced by the waveguide as given in Eqs. 12 (corresponding to the violet loop in (a)); i.e., the variation of $\gamma$ (nothing but $\Im(n)$) along both the $x$ and $z$ directions.

Now, we study the actual propagations of $\psi_0$ and $\psi_1$, while an EP is dynamically encircled along the length of the waveguide. For such a device level implementation of the convenient EP encircling scheme (as described in Eq. 6) dynamically, we have to realize the adiabatic simultaneous variations of $\gamma$ and $\tau$ along the closed loop by varying the imaginary part of the transverse index profile (as given in Eq. 3) along the propagation direction (i.e. $z$-axis) of the waveguide. To achieve adiabatic modes variation, we have to vary $\gamma$ and $\tau$ extremely slowly. Such a mapping on the waveguide exploiting paraxial approximation should obey the time-dependent Schrödinger equation with $z$ as the time axis. Here, the amount of loss ($\gamma$) is continuously tuned with proper maintenance of fractional loss ratio ($\tau$) in the two halves of the waveguide core along the $z$-direction; where for each transverse section (i.e. for each $z$-values), the index profile corresponds to a particular set of $(\gamma, \tau)$ lying on the parametric loop. This mapping is successfully achieved by considering $\phi = 2\pi z/L_0$ in Eqs. 6 as

$$\gamma(\phi) = \gamma_0 \sin\left(\frac{\pi}{L_0}z\right); \gamma_0 > \gamma_{EP}, \quad \tau(\phi) = \tau_{EP} - \eta \sin\phi \left(\frac{2\pi}{L_0}z\right). \tag{12}$$

Here, $L_0$ represent the total operating length (along z-axis) of the waveguide. For $0 \leq z \leq L_0$, above equations (Eqs. 12) describe the complete simultaneous variation of $\gamma$ and $\tau$ to enclose an EP along length of the waveguide; which is shown in Fig. 5(b). Now, we consider the input of the waveguide at $z = 0$ which indicates $\phi = 0$, and output of the waveguide at $z = L_0$ which gives $\phi = 2\pi$. This means one complete pass along the waveguide is equivalent to the one complete parametric loop around EP in the $(\gamma, \tau)$-plane (shown by the violet loop in Fig. 5). Thus, at both the input and output, $\gamma = 0$; where the supported modes are passive. Here, the clockwise and counter-clockwise encirclements along the closed parametric contour are realized by simply changing the direction of propagation along the waveguide.

Here, following the dynamical EP-encircling scheme along the length of the waveguide as described above, we numerically investigate the modal propagations. As the variations of $\Im[n]$ along the z-direction are extremely slow (usually, quasi-static) in comparison with the order of the wavelength ($\lambda = 2\pi$; as we normalize $\omega = 1$), the 2D scalar wave equation can be reconstructed with the paraxial approximation as

$$2i\omega\partial_z\psi(x,z) = -[\partial_x^2 + \Delta n^2(x,z)\omega^2]\psi(x,z) \tag{13}$$

Here, $\Delta n^2(x,z) \equiv n^2(x,z) - n_l^2$. We exploit the split-step-Fourier method to solve the Eq. 13 [30].

In Fig. 6 we show the corresponding beam propagation simulations results. Here, we judiciously choose the total device length $L_0 = 5 \times 10^3$ in dimensionless unit (equivalent to $\approx 796$ free-space wavelengths). At first, we consider the clockwise dynamical encirclement around EP by setting $\eta = -0.05$. To realize this operation in our waveguide, we launch the eigenmodes individually at the input i.e., at $z = 0$; where both of them are initially free of loss. In the Fig. 6(a.1) and (a.2), we depict the propagations of $\psi_0$ and $\psi_1$; where it is clearly observed that both of them are effectively converted to $\psi_1$ at the output $z = L_0$. This conversion is irrespective of the choice of the input modes. Here $\psi_0$ evolves adiabatically and converted into $\psi_1$; however, $\psi_1$ behaves non-adiabatically and return to itself after end of the encircling process. Now, to investigate the effect of the anticlockwise dynamical encirclement around the EP (setting $\eta = 0.05$), we just change the direction of propagation and launch the eigenmodes at the output end i.e., at $z = L_0$. Corresponding propagations of $\psi_0$ and $\psi_1$ are shown in Fig. 6(b.1) and (b.2) respectively. Interestingly, in this case, $\psi_0$ evolves non-adiabatically, whereas $\psi_1$ evolves adiabatically and converted into $\psi_0$. Thus, at the input $z = 0$, we retain $\psi_0$, regardless of the choice of launched mode at $z = L_0$.

So, $\psi_0$ and $\psi_1$ experience different decay rates; which are calculated by averaging the losses over the entire contour by

$$\gamma_{av} = \frac{1}{2\pi}\int_0^{2\pi} \Im(\beta)d\phi \tag{14}$$

(considering the adiabatic expectation of $\beta$-s as shown in Fig. 4(b)). Now, we calculate the $\gamma_{av}$ for red trajectory as 0.167; however, for blue trajectory 0.176. Now, along the red trajectory $\beta_0$ evolves and going to the initial position of $\beta_1$. Thus, for clockwise dynamical encirclement around the EP, $\psi_0$ propagates with lower decaying rate and we get the conversions $\psi_0 \to \psi_1$ and $\psi_1 \to \psi_1$; as expected after the non-adiabatic corrections. Similarly, it can be observed that for anticlockwise encirclement around the EP, $\beta_1$ evolves with lower loss and going to the initial position of $\beta_0$; which gives $\psi_1 \to \psi_0$ and $\psi_0 \to \psi_0$. Thus, Fig. 6 represents an efficient chirality drive asymmetric mode conversion; where only the choice of direction with which the embedded EP is encircled regulates whether the output mode is $\psi_0$ or $\psi_1$, irrespective of the choice of input excited mode.

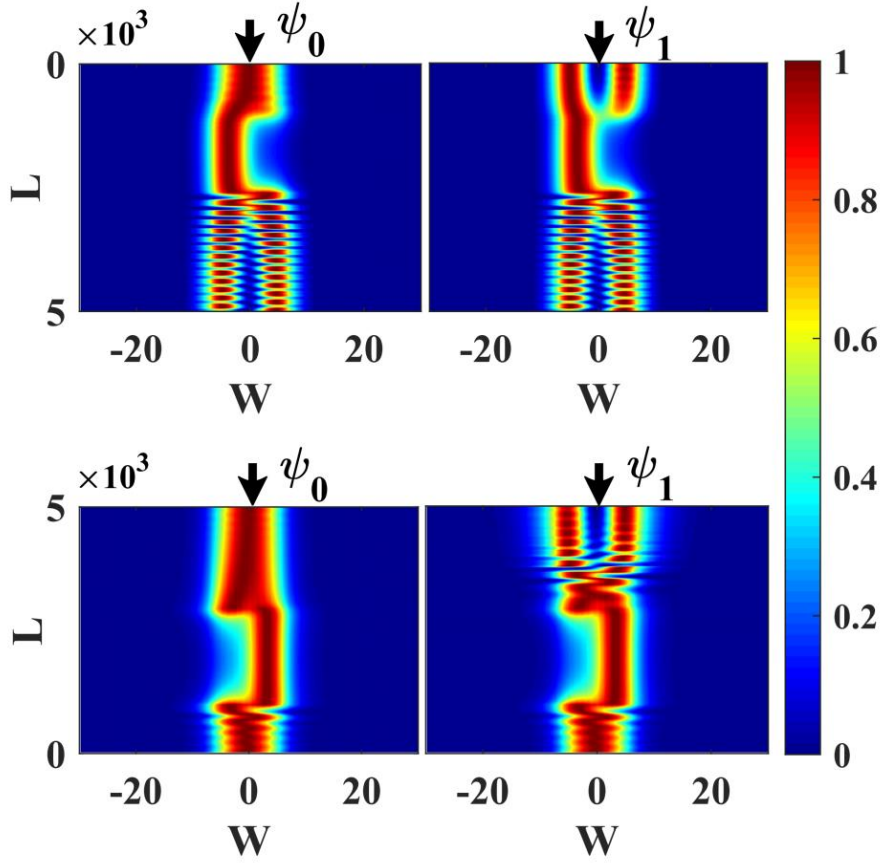

**FIG. 6.** (Color online) *Beam propagation simulation results*. **(a.1)** Adiabatic conversion from $\psi_0$ to $\psi_1$ and **(a.2)** nonadiabatic evolution of $\psi_1$ ($\to \psi_1$) for clockwise dynamical encirclement around the EP; where both the $\psi_0$ and $\psi_1$ are launched at the input ($z = 0$) and at the output ($z = L_0$) they are converted into $\psi_1$. **(b.1)** Non-adiabatic evolution from $\psi_0$ ($\to \psi_0$) and **(b.2)** adiabatic conversion of $\psi_1$ to $\psi_0$ for anticlockwise dynamical encirclement around the EP; where both the $\psi_0$ and $\psi_1$ are launched at the output ($z = L0$) and at the input ($z = 0$) they are effectively converted into $\psi_0$. For proper visualization, we re-normalize the intensities (shown by color-bar) for each $z$, so the overall intensity change is not displayed.

In Fig. 7(a) we plot the input normalized squared modal fields i.e., $|\psi(x)|^2$. Fig. 7(b) and (c) shows the output fields when the EP is dynamically encircled along the clockwise and anticlockwise directions respectively. The intensities of the input and output fields are shown at the waveguide ends i.e., either at $z = 0$ or at $z = L0$; when they are free of loss. Fig. 6 and 7 reveal that our designed waveguide, operating at an EP, can be exploited as an efficient chirality driven optical mode converter; where one can reach the desired output by changing the direction of light propagation with proper dynamical distribution of only loss-coefficient following the encircling contour around the EP.

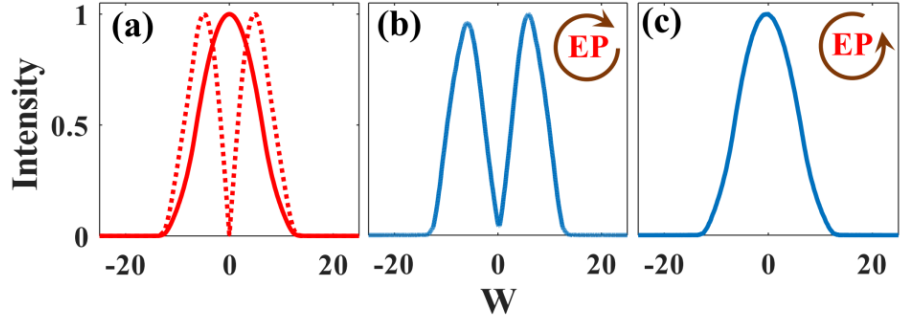

**FIG. 7.** (Color online) *Chirality driven mode conversion.* **(a)** Input field intensities of $\psi_0$ (red solid curve) and $\psi_1$ (red dotted curve). **(b)** Effective output field intensity at $z = L_0$ when the input eigenmodes are launched from $z = 0$ (for clockwise EP encirclement process). **(c)** Effective output field intensity at $z = 0$ when the input eigenmodes are launched from $z = L_0$ (for anticlockwise EP encirclement process).

We also calculate the conversion efficiencies during the EP-aided asymmetric mode conversion phenomena exploiting the overlap integrals between the input and output fields as

$$C_{ij} = \frac{\left|\int \psi_i \psi_j dx\right|^2}{\int |\psi_j|^2 dx \int |\psi_j|^2 dx} \qquad i,j = 0,1; \; i \neq j \tag{15}$$

Here, $C_{ij}$ defines the conversion efficiency from the $i^{th}$ mode to the $j^{th}$ mode. Now, for the described mode conversion phenomena in Fig. 6, we calculate the conversion efficiencies for the conversions $\psi_0 \to \psi_1$ and $\psi_1 \to \psi_1$ as 95.32% and 92.93% respectively; while the EP is encircled along the clockwise direction. However, when we encircle the EP along the anticlockwise direction, the efficiencies for the conversions $\psi_1 \to \psi_0$ and $\psi_0 \to \psi_0$ are 70.64% and 69.86%, respectively.

To take into account the fabrication tolerances during devise-based applications of such EP-aided mode conversion phenomena, we append some deliberate random fluctuations up to 10% (approximately) on the both $\gamma$ and $\tau$ when they are dynamically varied along the closed contour (described by Eq. 12) and recheck the propagations of $\psi_0$ and $\psi_1$ along the waveguide. Here, it is observed that asymmetric mode conversion phenomenon is omni-present and robust against the parametric fluctuations. For clockwise operation, the conversion efficiencies differ by $\approx 0.83\%$ (average); whereas for anticlockwise operation the conversion efficiencies differ by $\approx 1.97\%$ (average) from the previously calculated values.

Note that, as we have considered an all-lossy structure, the eigenmodes should be leaky. Thus, during this asymmetric mode conversion phenomenon, a crucial limiting factor is the decay length of the eigenmodes. Here, the choice of effective operating length of the waveguide $L_0$ is important. For large $L_0$, the excited eigenmodes can be decayed before conversion; which is not desirable. To overcome this limitation, the conventional time reverse method that usually valid for $PT$-systems (for which modes are amplifying) should not be exploited as our all-lossy waveguide structure is not time symmetric. Thus, for efficient device applications, concern should be taken to choose $L_0$, which should be shorter than the decay length of both eigenmodes.

## 5. CONCLUSION:

In summary, we exclusively report a step-index symmetric planar all-lossy optical waveguide which supports only two modes. Two leaky modes are coupled and analytically connected by a second order exceptional point (EP); where the interactions between them are controlled by tailoring the transverse distribution of attenuation profile only. Here, the overall loss profile is characterized by two parameters viz. loss-coefficient ($\gamma$), and the fractional-loss-ratio ($\tau$); especially which maintains a specific loss difference between two halves inside the core region. Such an all-lossy waveguide structure, operating at an EP, is reported for the first time to the best of our knowledge; where no external gain is present. Enclosing an embedded EP in parameter plane properly (by two different types of enclosing contours), the characteristics of modal propagation constants ($\beta$) around the EP are investigated; where it is evident that for one complete encirclement, the $\beta$-values corresponding to the pair of coupled modes adiabatically exchange their identities and switch between each other. Finally, we present an impactful study on the propagations of the supported eigenmodes, when the EP is dynamically encircled with simultaneous quasi-static variation of the system parameters along the length of the waveguide. Here, due to breakdown in adiabaticity, we observe an efficient asymmetric mode conversion phenomenon which is governed by the direction only along which the respective EP is encircled. When the light propagates through the waveguide then due to effect of the EP, it is converted into a specific mode, irrespective of the choice of the input mode. The effective output mode is dictated by the direction of light propagation along which the loss modulation dynamically encircles the EP in either clockwise or anticlockwise direction. This all-lossy waveguide structure enriched with the exclusive scheme of chirality driven asymmetric-mode conversion can be extensively implemented on state-of-the-art integrated on-chip device-based applications like all-optical on-chip mode converters, circulators, filters, etc.

## ACKNOWLEDGMENTS:

AL and SG acknowledge the financial support from the Science and Engineering research Board (SERB) under Early Career Research Grant [ECR/2017/000491], and the Department of Science and Technology (DST) under INSPIRE Faculty Fellow Grant [IFA-12, PH-23], Ministry of Science and Technology, India.

## REFERENCES:

1. N. Moiseyev, *Non-Hermitian Quantum Mechanics* (Cambridge University Press, Cambridge ; New York, 2011).
2. T. Kato, *Perturbation Theory for Linear Operators* (Springer, Berlin, 1995).
3. W. D. Heiss, and A. L. Sannino, J. Phys. A: Math. Theor. **23**, 1167 (1990); W. D. Heiss, Phys. Rev. E **61**, 929 (2000).
4. J. Doppler, A. A. Mailybaev, J. Böhm, U. Kuhl, A. Girschik, F. Libisch, T. J. Milburn, P. Rabl, N Moiseyev, and S. Rotter, Nature **537**, 76 (2016).
5. G. Chen, R. Zhang, and J. Sun, Sci. Rep. **5**, 10346 (2015).
6. X.-L. Zhang, S. Wang, B. Hou, and C. T. Chan, Phys. Rev. X **8**, 021066 (2018).
7. S. Ghosh, and Y. D. Chong, Sci. Rep. **6**, 19837 (2016).

8. A. Laha, and S. Ghosh, J. Opt. Soc. Am. B **34**, 238 (2017); A. Laha, A. Biswas, and S. Ghosh, , J. Opt. Soc. Am. B **34**, 2050 (2017).
9. Y. D. Chong, L. Ge, and A. D. Stone, Phys. Rev. Lett. **106**, 093902 (2011).
10. M. Liertzer, L. Ge, A. Cerjan, A. D. Stone, H. E. Tureci, and S. Rotter, Phys. Rev. Lett. **108**, 173901 (2012).
11. K. Ding, Z. Q. Zhang, and C. T. Chan, Phys. Rev. B **92**, 235310 (2015).
12. D. A. Bykov and L. L. Doskolovich, Phys. Rev. A **97**, 013846 (2018).
13. R. Lefebvre, O. Atabek, M. Šindelka, and N. Moiseyev, Phys. Rev. Lett. **103**, 123003 (2009).
14. H. Cartarius, J. Main, and G. Wunner, Phys. Rev. Lett. **9**, 173003 (2007); Phys. Rev. A **79**, 053408 (2009).
15. B. Dietz, H. L. Harney, O. N. Kirillov, M. Miski-Oglu, A. Richter, and F. Schafer, Phys. Rev. Lett. **106**, 150403 (2011).
16. C. Dembowski, H-D. Graf, H. L. Harney, A. Heine, W. D. Heiss, H. Rehfeld, and A. Richter, Phys. Rev. Lett. **86**, 787 (2001).
17. C. Dembowski, B. Dietz, H-D. Gr¨af, H. L. Harney, A. Heine, W. D. Heiss, and A. Richter A, Phys. Rev. E **69**, 56216 (2004); Phys. Rev. Lett. **90**, 034101 (2003).
18. L. Feng, X. Zhu, S. Yang, H. Zhu, P. Zhang, X. Yin, Y. Wang, and X. Zhang, Opt. Exp. **22**, 1760 (2014)
19. C.-H. Yi, J. Kullig, and J. Wiersig, Phys. Rev. Lett. **120**, 093902 (2018).
20. H. Hodaei, A. U. Hassan, W. E. Haynga, M. A. Miri, D. N. Christodoulides, and M. Khajavikhan, Opt. Lett. **41**, 3049 (2016).
21. W. Chen, S¸. K. Özdemir, G. Zhao, J. Wiersig, and L. Yang, Nature **548**, 192 (2017).
22. H. Hodaei, A. U. Hassan, S. Wittek, H. G-Gracia, R. ElGanainy, D. N. Christodoulides, and M. Khajavikhan, Nature **548**, 187 (2017).
23. J. Wiersig, Phys. Rev. Lett. **112**, 203901 (2014); Phys. Rev. A **93**, 033809 (2016).
24. Y. Choi, C. Hahn, J. W. Yoon, S. H. Song, and P. Berini, Nat. Comms. **8**, 14154 (2017).
25. S.-Y. Lee, J.-W. Ryu, S. W. Kim, and Y. Chung, Phys. Rev. A **85**, 064103 (2012).
26. R. Uzdin, and N. Moiseyev, Phys. Rev. A **85**, 031804 (2012); R. Uzdin, A. Mailybaev, and N. Moiseyev, J. Phys. A: Math. Theor. **44**, 435302 (2011).
27. I. Gilary, A. A. Mailybaev, and N. Moiseyev, Phys. Rev. A **88**, 010102(R) (2013); I. Gilary, and N. Moiseyev, J. Phys. B: At. Mol. Opt. Phys. **45**, 051002 (2012).
28. T. J. Milburn, J. Doppler, C. A. Holmes, S. Portolan, S. Rotter, and P. Rabl, Phys. Rev. A, **92**, 052124 (2015).
29. S. Phang, A. Vukovic, S. C. Creagh, T. M. Benson, P. D. Sewell, and G. Gradoni, Opt. Express **23**, 11493 (2015).
30. G. Agrawal, *Nonlinear Fiber Optics 5th ed.* (Academic Press, USA, 2012).